# On the role of ethics and sustainability in business innovation

Maria Fay[1,2], Frederik F. Flöther[3]
[1] University of Münster, Schlossplatz 2, 48149 Münster, Germany
[2] SAP Switzerland AG, The Circle 66, 8058 Zürich, Switzerland
[3] QuantumBasel, Schorenweg 44b, 4144 Arlesheim, Switzerland

## Abstract

For organizations to survive and flourish in the long term, innovation and novelty must be continually introduced, which is particularly true in today's rapidly changing world. This raises a variety of ethical and sustainability considerations that seldom receive the attention they deserve. Existing innovation adoption frameworks often focus on technological, organizational, environmental, and social factors impacting adoption. In this chapter, we explore the ethical and sustainability angles, particularly as they relate to emerging technologies, artificial intelligence (AI) being a prominent example. We consider how to facilitate the development and cultivation of innovation cultures in organizations, including budding startups as well as established enterprises, through approaches such as systems thinking.

## Background

Analysis of the Fortune 500 list over time shows that new companies and business models have been rapidly emerging in recent years[1]. Those companies on the list that have already existed for several decades have generally undergone multiple fundamental shifts in business area and strategy, for instance IBM and Toyota. Hence, it is imperative that organizations continually introduce new ideas, methods, and even ways of thinking in order to stay competitive and relevant. This applies to established organizations as well as startups and new institutions.

Developing and maintaining such an influx of novelty raises many questions that demand careful analysis. Innovation adoption frameworks have been developed, such as the TOE framework[2], which cover technological, organizational, environmental, and in some cases even social factors. Topics that have been addressed in these categories respectively include, for example:
- Security
- Enablement
- Regulation
- Trust

There has also been some work around exploring the connections between ethics and innovation. For example, in considering the relationship between ethics and innovation it was argued that they are much more entangled and have a much greater degree of overlap, in a positive rather than a conflicting sense, than might seem to be the case from a first glance[3]. In addition, a theoretical model was developed for the governance of innovation processes, and it was argued that companies will intrinsically consider ethical aspects as they build long-term strategies[4]. While the mechanics and optimal processes to engender business innovation have been considered[5], inclusivity has been explored in this context[6], and innovation ethics have been studied in conjunction with international development[7], a broader exploration of the ethics associated with innovation policies is required.

Furthermore, in recent years there has been a surge of global awareness of and interest in sustainability. Here, business sustainability is defined to be doing business without negatively impacting the environment, community, or society as a whole[8]. Sustainability and ethics have close ties, the former being grounded in ethical commitment to the wellbeing of both current and future generations[9].

However, extensions of frameworks around technology adoption in organizations with ethical as well as sustainability dimensions have not yet been deeply explored. This is the focus of the present chapter. The remainder of the text is structured as follows. First, pertinent ethical considerations for the development of novelty in organizations are discussed. Second, light is cast upon the variegated role of sustainability in organizational innovation processes. Third, recommendations are formulated with regard to the inclusion of ethics (E) and sustainability (S) dimensions in technology adoption and business innovation beyond the traditional aspects of technology (T), organization (O), and the environment (E), thus resulting in a "TOEES" framework. Finally, conclusions are drawn.

**Ethical considerations**

Introducing, or even just fostering, innovation in an organization comes with priority setting, opportunity costs, and (often) changes. Therefore, values and ethical dimensions have to be elucidated[10]. These do not have to be thought of as "brakes" to innovation; on the contrary, they may be conducive to novel ideas[11] and represent great opportunities with regard to product, service, and organizational innovation[12]. Given their varied nature, it is not possible to consider every possible dimension and situation. In this section, a number of quite different scenarios are discussed in order to illustrate the range and complexity of the questions.

*Space and time for innovation*

Consider a company that wants to nurture entrepreneurship and innovation culture. Often good ideas come through impromptu face-to-face interactions among people, at the (proverbial) water cooler. A straightforward way to increase the frequency and impact of such interactions is creating an environment that promotes exchange and creativity, provides comfort, and takes many of the daily "problems" off employees' shoulders (free meals, gym access, evening events, …). This is how many companies have proceeded, Silicon Valley campuses being among the pioneers.

Further consider a company's headquarters where every single need is met, from supermarkets to childcare facilities to cinemas to accommodation. Taking the idea this far, not unlike the campus in the (in)famous "The Circle" novel/movie, the center of employees' lives are effectively relocated to the workplace. Such a shift is even quantifiable, the number of hours per week spent by employees on the campus being a simple measure.

Is this ethical? It very much depends on a given employee. Some may welcome such an environment with open arms. Others might prefer more time "at home" due to the nature of their tasks or due to personal reasons, particularly given the recent shift towards hybrid work models[13]. It is also a matter of a given company's expectations towards its employees, as peer pressure and culture in general make a great difference.

*Work product – a product of…work?*

What is "work", when employees are so dedicated to what they do that they do not treat it as work and willingly spend free time on those initiatives? What if one's hobbies result in a work product? Most employment contracts regulate the results of work in the field where a person is employed as well as the reporting of side activities – which all makes perfect sense to avoid conflicts of interest and intellectual property (IP) rights infringement. Any organization would like to hire professionals who are experts in their respective fields, are intrinsically motivated, and are constantly developing and learning. On the way to leveraging and developing their expertise, those employees might want to teach a course at a university, become a mentor, present keynotes, or write a book. Given that role and company changes are becoming more frequent, it is becoming more difficult to determine if a given piece of thought leadership is the product wholly or partially due to an individual's time in a certain role. How will IP laws change in the future with AI coming into play?

*Human response to (technological) changes and evolution of jobs*

Naturally, changes in the environment typically lead to changes in the feelings and actions of humans. Changes in technology are no exception[14], and people may welcome or dread them, perhaps even both at the same time to a degree; medical innovation is a case in point[15]. On the one hand, new technology may lead to novel tools that one can play with and which perhaps even make work easier. This goes hand in hand with learning opportunities and chances for stimulating interactions with others who already know about or perhaps are also learning about the new tool.

On the other hand, technological changes generally engender drastic changes in the job market, reducing demand for certain roles while increasing demand for other occupations. Thus, one must keep the question of how introduction of new technology affects the workforce always top of mind. There is early evidence that digitalization leads to increased employment of high-skilled workers and reduced employment of low-skilled workers[16]. There are entirely new job types which are required, prompt engineers being one example[17]. Given the pace of technological development (as well as the abstractness of, for instance, quantum technology[18]), all workers should be open, and ideally intrinsically motivated, as concerns life-long learning. As such, introduction of new ideas, methods, and technology as well as general business innovation should go hand in hand with fostering a culture of enablement and learning.

*Example of artificial intelligence (AI)*

Consider the rise of AI. After decades of research, AI has now achieved mass awareness as a result of various breakthroughs that have been enabled by factors such as greater data availability, enhanced computational power, and improved algorithms. While movie-like human-like systems and robots are likely still quite some way off, artificial general intelligence (AGI), that is general-purpose systems approaching or exceeding human intelligence[19], now seems no longer quite such an illusory prospect. Sparks of AGI are arguably already seen in today's advanced generative AI systems[20]. This has induced public debate, and the EU AI Act represents one of the first regulatory frameworks around AI[21].

The ethical questions surrounding AI, in general and in the context of business innovation, are manifold. For instance, consider creative work. Previously, it was (relatively) clear that novel output was the result of one or more humans. Today, an engaging document, image, audio file, or video could be the result of machine or human-machine processes. In both cases, systems were likely trained with data sets from diverse sources. As such, many people contribute, via their data which is used to train new AI systems, incrementally to many novel outputs. Clearly, this raises difficult ethical questions as to how such outputs may be used and attributed and who should be rewarded for successes. What contribution percentage belongs to the people sharing the data, the people designing the systems, the people providing generative AI prompts, the people leveraging the raw system outputs… and the system itself? There is an evident risk of "responsibility gaps"[22]. To complicate (ethical) matters further, some of these outputs may be of great interest to the public, a case in point being medical innovation.

Another area where AI impacts business innovation ethics is in democratized access. AI, and other emerging technologies such as quantum computing, are believed to have revolutionary potential; as such, a global race has started between organizations and countries in order to be the first to fully develop and take advantage of these technologies[23]. Should early-mover organizations and countries be permitted to reap all these benefits as a result of their likely significant investments? Or is some form of democratized access, leading to a more level playing field, a better ethical approach?

## Sustainability considerations

Sustainability and innovation are closely interlinked, and more and more companies acknowledge that. This includes not only bringing in sustainability aspects into new products (by making them more circular or reducing emissions) but also creating new roles. Holcim – a Swiss building materials giant – has introduced the role of Chief Sustainability and Innovation Officer (CSIO) in 2021[24]. In one of her keynotes in 2022, Magalie Anderson (Holcim's CSIO at the time) explained that this allows the company to truly introduce sustainability into every product and process, from sustainable innovation in design and R&D to production to recycling. The achievements have been impressive: in 2022, Holcim became the first company in its market segment to achieve a "double A" rating from the Climate Disclosure Project in both climate and water and has been driving social efforts with technology such as "building" the world's first 3D-printed school in Malawi[25]. Therefore, it seems that linking sustainability and innovation also on the organizational level drives clear benefits.

Furthermore, it is of utmost importance to consider sustainability holistically. There is a reason the United Nations (UN) has chosen 17 sustainable development goals (SDGs)[26] as they are addressing various aspects of how development can be sustainable. They range from improving education to ensuring access to healthcare and clean water to preserving biodiversity. A new product – such as 1-dollar glasses[27] – might help address poverty and improve healthcare; still, its production and recycling will have a certain carbon footprint, consuming water and energy. Taking a holistic perspective, would this product be sustainable?

Based on such considerations and given that sustainability is very much multifaceted, it is necessary to illuminate the intersection of sustainability and technology adoption from different angles. Below are several sustainability dimensions that play an important role in an innovation context.

*Sustainability impact of innovation*

The phrasing of this subtitle deliberately highlights the "sustainability" impact without referencing a specific environmental or product carbon footprint due to the reasons listed above. When considering innovation and new ideas, it is important, for instance, to estimate to what degree a novel product would be more sustainable than previous ones, and, in addition, what impact a changed process could create. There are multiple methods allowing calculations of this type to be carried out, such as a lifecycle assessment (LCA) or product carbon footprint measurement. Still, hardly any take a holistic approach and go beyond climate or environmental effects, for example considering social impact or inequality reduction aspects, biodiversity changes, and responsible consumption. While there are alternative approaches, such as social and environmental life cycle assessments (S-LCA, E-LCA)[28] for assessing the contributions of products to the 61 out of 169 targets set by the UN SDGs, the feasibility of scaling implementation on the product level at every company can be rightfully questioned.

In the end, a lot comes down to what the companies that manufacture those products and create innovation are driven by. They are all part of a global market economy, and they have to comply with the market mechanisms that exist as well as regulations. But what if every company and every decision-maker would go just one step further? Perhaps they could come up with a recycling process for the products which they produce or try to reduce their biodiversity footprint (even if it is not yet required by law). Reflecting on what has been said earlier in this chapter, it is a question of ethics and of personal and company values. Is it ethical "just" to do what you were hired to do – or is it more ethical to go the extra mile? Is it ethical to expect extra miles from others?

*Innovation for sustainability*

Creative products and ideas meanwhile contribute significantly to sustainability progress. Consider the following perspectives and examples.

"Technology innovation lens"

It is no coincidence that "climate tech" and "green tech" are two terms that have been trending in the last few years. Every industry has thousands of brilliant innovation examples that contribute to sustainability, especially through research and technology improvements.

A lot is being done on the forefront of materials research, from biodegradable recycled plastics for consumer products and textiles to mushroom growth for fabric production and as an alternative to carbon-intensive concrete, plastic, and foam in building insulation. Another example is food technology that can make an impactful contribution by addressing malnutrition and hunger. It is crucial to make advances and scale alternative proteins, develop precision fermentation along with other technologies, and enable more communities to have reliable access to food.

Finally, energy technology and renewables represent a rapidly growing area. Studying, for instance, the list of global climate technology unicorns[29] shows that many energy and mobility companies are represented in the top 20, with Northvolt, a battery technology company, being the frontrunner. Clearly, whether it is about designing new batteries or meals, technological breakthroughs are key enablers on the road towards renewable materials and sustainable processes.

"Circular economy lens"

Innovation does not always manifest itself in new products. The circular economy is transitioning from traditional linear "take-make-waste" thinking to closing the loop by extending product lifecycles through reuse, refurbishment, and remanufacturing. Hence, in the case of remanufacturing, when a product comes to its end of life (for instance, a mobile phone that no longer works because its battery is dead), one returns it to the original manufacturer or recycler. They then dismantle the product and take out valuable components that could still be reused for repair or remanufacturing ("the process of bringing a used product to like-new condition through replacing and rebuilding component parts"[30]). Is this innovative? Well, it requires a new process (such as product takeback and its quality verification) with new skills and tools related to it, new design guidelines (and sometimes changing product specifications), and further updates. Is this sustainable? It allows one to reduce (even avoid) raw material (as well as water and energy) consumption and waste. As a result, this constitutes yet another example of innovation for sustainability.

Furthermore, the circular economy is also about refusing (so not producing or buying in the first place), rethinking, and reducing consumption[31]. As an example, consider a refillable water bottle that one carries to work or the gym or while traveling. In a way, this is a small piece of innovation introduced into one's daily routine / drinking process, which allows one to refuse using plastic water glasses, rethink consumption, and perhaps keep the bottle for a long time (even more so, if personal meaning is attached to it).

"Frugal innovation lens"

Frugal innovation is a perfect example of how new ideas and products can address sustainability on multiple levels. One of the classical definitions and key premises for frugal innovation is "achieving more with less"[32]. This pertains to products that are simple and affordable, address real needs, and are easy to produce and use. Consider the problem of early breast cancer screening in developing countries. Traditional mammography equipment (large, expensive) is usually available at bigger hospitals. Thus, living in a rural area, often one cannot afford to travel to a hospital just for a checkup and needs to wait for a long time to get an appointment. This leads to the fact that most women in such areas do not undergo prophylactic examinations until it is too late. What if there were an easy-to-use-and-produce handheld device that could be straightforwardly brought to rural areas and employed for regular screenings of women living there to catch early signs of disease?

This was the idea by UE LifeSciences, a company offering iBreastExam[33]. Company representatives travel to distant areas to train more people on using this simple hand-held system. Although the device cannot achieve the precision of professional mammography equipment, it can detect most knots (even small ones). In case a malicious knot is suspected, a woman can then travel to a larger hospital for further testing. Therefore, just one such novel idea can address a tremendous healthcare problem for large populations, especially in areas with great poverty – and therefore make a significant contribution to sustainable development. Such simple and scalable solutions could become major enablers across the globe as the world attempts to race towards maintaining a suitable climate.

"Ecosystem innovation lens"

Sustainability demands cooperation from various parties. One cannot calculate product footprints alone: for this, one needs data from one's suppliers and partners for Scope 3 emissions (which represent emissions that an organization only indirectly affects in its value chain).

How is collaboration becoming relevant in the innovation context? First, it is about jointly defining problems and developing solutions. For instance, this is what the Catena-X[34] automotive network is doing on the industry level. Such initiatives can help ensure acceptance, adoption, and ease of use by relying on common standards and platforms. Second, crowdsourcing and crowdfunding are powerful. There are many great ideas and businesses whose first steps happened on Kickstarter or similar platforms – and early feedback together with other advantages such platforms offer is a crucial element of creating a sustainable product. Third, even without being directly connected to others one can be a part of sustainability ecosystems and exchange ideas. Consider projects such as Local Happinez. This is a platform that allows users to find and support a local sustainable initiative of their choice next time they are on vacation[35]. Such a solution is simple but effective, and there are both direct and indirect benefits to it. A direct benefit is what the goal of the project is and what the community gets out of it: a cultivated coral reef or renovated water pipelines or harvested crop. One should not underestimate the indirect benefits, which include what participants in those projects learn from one another, making them more conscious of the world's ecosystem and the impact of decisions they make every day. Therefore, such projects have a strong educational effect, contributing to sustainability awareness, and thus also ethical dimensions, altogether.

*Sustainability of innovation lifecycles*

A first question arises: *when is it time for innovation?* When is it time for existing methods and ideas to be replaced by more novel ones? This taps into psychology and philosophy and is closely interlinked with management practices; it applies not only to product novelty but also to practices and processes. Consider company reorganizations. How ready are employees for such changes? How many are wondering why it is time again for a change if new structures are not yet in place from previous changes? Still, often companies fail to change in time, for example by not opening a new division to design an original product, and instead elect to focus on established practices instead. The concept of organizational unlearning[36] is helpful here to understand how and when organizations need to "forget" old practices. By creating situational awareness, providing temporal and spatial freedom, encouraging an error-forgiving culture, and reducing the influence of old knowledge over time, companies can create room for new knowledge and learning[37]. In this way, the spirit of innovation can be sustained and flourish.

Now the next question comes: *how can one ensure that innovation persists?* New technology and methods should not just be implemented but also accepted, scaled, and further improved. While multiple frameworks around innovation management exist, one that has earned its reputation in the field of technology innovation is highlighted here (future business innovation will likely often be at least partially driven by technology innovation). This is the technology-organization-environment (TOE) framework, as originally proposed in 1990[38] and as employed in later applications and adaptations[39]. It defined three key dimensions (technology, organization, environment) that have impact on technology adoption and scalability. For instance, within a company, its size and structure as well as communication processes and slack built over time (c.f. the aforementioned organizational unlearning) influence the speed and success of technology adoption. In terms of the external environment (the focus here lies on the market and policies rather than climate), market structure or government regulations could also be of major importance. A prominent recent example is the historic EU AI Act[40], which regulates development and usage of AI and thus impacts a great many startups and companies across industries. Linking this with the original question around sustaining innovation in an organization, one needs to navigate between all those dimensions and stay on top of developments in a given company and in the market, which is becoming more and more difficult over time. Ironically, AI itself may also help overcome this barrier[41].

## Recommendations

Ethics and sustainability need to become integral parts of frameworks used in innovation and technology management. Taking the TOE example further, while TOE includes the "environment" dimension, this refers to contextual information about the external task environment and not (explicitly) to sustainability angles. There is therefore a need for more holistic approaches which, building on the earlier considerations from this chapter, include the dimensions of ethics and sustainability. Figure 1 illustrates an adaptation of the TOE framework which has been extended with ethics and sustainability ideas, thus becoming the "TOEES" framework.

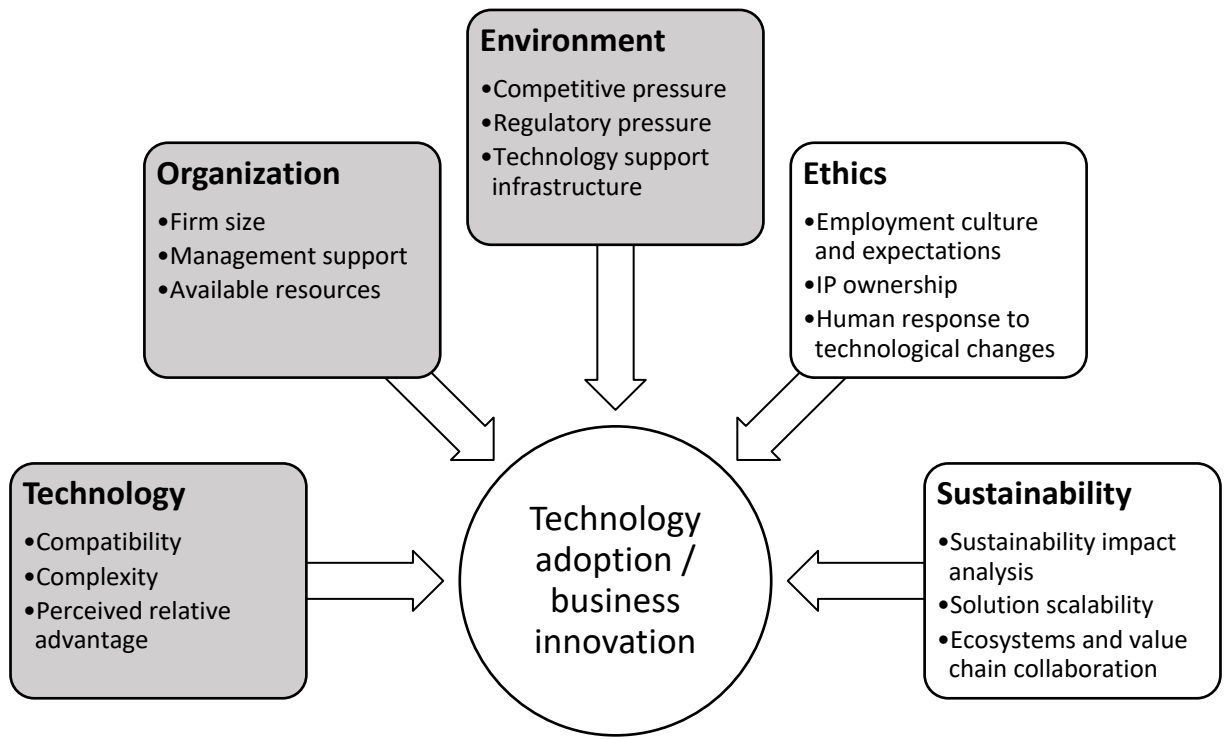

Figure 1: Adapted TOE framework[42] that has been extended to explicitly include the dimensions of ethics and sustainability, thereby forming the "TOEES" framework.

Furthermore, systems thinking is crucial to ensure holistic understanding of innovation and its impact; responsible innovation requires exploration of individual, organizational, and system levels[43]. System dynamics as a discipline has been around for decades[44]; given the increasingly interconnected and complex world, it is further gaining in relevance. In 1972 the "Limits to growth" report[45] utilized a computer model to simulate the consequences of interactions between the Earth and human systems. Moreover, systems thinking has been described as "a framework for seeing interrelationships rather than things, for seeing patterns rather than static snapshots"[46]. It is extremely difficult to account for all the possible consequences of decisions that one makes and may consider ethical or the innovation one introduces and how this influence sustainability in a broader sense. Modeling those interactions in real life might not always be realistic or even possible; however, reflecting on the points outlined in this chapter can help explore at least some of the pertinent considerations.

Consider the case of Amazon Echo in terms of an anatomical map that requires human labor, data, and planetary resources for the unit to function[47]. In turn, such a unit, powered by AI, becomes part of a great breadth of decisions and processes; as with any technology, it may help with "good" or "bad" acts[48]. Clearly, the ethics and sustainability of a new product such as Amazon Echo requires a broad analysis and, hence, systems thinking is very apt.

Finally, leveraging communities and ecosystems is key when it comes to ethical and sustainable business innovation. Sustainability is impossible to achieve alone: one must have suppliers on board who follow the principles of ethical sourcing and reduce the environmental footprint of their products. One also has to have consumers on board who "vote" for one's sustainable product with their money. Thus, one needs partners, including academia, enterprises, startups, and the crowd, to make even more impactful innovation happen.

**Conclusion**

Ultimately, technology adoption and business innovation are highly complex, depending on many aspects that may also change over time. The rise and fall of companies, including the degree to which they are able to remain at the forefront of innovation, are testament to this. Particularly in recent years, many ethical and sustainability dimensions have gained importance in this regard. Today, they must be considered as businesses find their footing in an age with hybrid work models, accelerating AI, and mandates necessitating sustainable business models embedded in circular economies. These changes take time to internalize. Awareness and discourse thus represent a sensible first step; the "TOEES" framework outlined in this chapter is meant as an enabler in this regard. Clearly, this must then be complemented with concrete actions and plans, which may leverage systems thinking. Only organizations that apply the latest ideas to remain flexible, nimble, and proactive in how they treat business innovation… will (continue to) be successful in achieving business innovation.

**About the authors**

Dr. Maria Fay is a Sustainability & AI Director at SAP Switzerland AG[1] and a guest researcher at the University of Münster. With over a decade of experience in business and technology consulting, Maria has been advising companies around the world on the adoption of emerging technologies, most recently focusing on AI. She holds degrees in business economics, informatics, and sustainability, and has also lectured on digital business models. She is a host of the "Apropos… Nachhaltigkeit" sustainability podcast. Moreover, Maria is an advisory board member for The Inclusion Foundation and engages in mentoring and volunteering work. She has worked and lived in Switzerland, Liechtenstein, Germany, and Russia.

**Name:** Dr. Maria Fay
**Position:** Guest Researcher
**Affiliation:** University of Münster
**Address:** Leonardo Campus 3, 48149 Münster
**Email:** maria.fay@uni-muenster.de
**ORCID:** 0000-0003-3626-0413

Dr. Frederik F. Flöther is Chief Quantum Officer at QuantumBasel, the first commercial quantum hub of Switzerland. Prior to that, he was over 7 years with IBM, most recently responsible for healthcare and life sciences at IBM Quantum. Frederik researches the technological applications as well as philosophical implications of quantum computing and AI. He was elected to the IBM Academy of Technology and appointed Master Inventor and Qiskit Advocate. Frederik holds a PhD in physics, with a focus on photonic quantum computing, and a MA, MSci, and BA from the University of Cambridge as well as a professional certificate in quantum computing applications from MIT. In total, he has authored over 40 filed patents, peer-reviewed publications, white papers, and book chapters. Frederik has lived/worked in Switzerland, USA, United Kingdom, Belgium, and Germany.

**Name:** Dr. Frederik F. Flöther
**Position:** Chief Quantum Officer
**Affiliation:** QuantumBasel
**Address:** Schorenweg 44b, 4144 Arlesheim, Switzerland
**Email:** frederik.floether@quantumbasel.com
**ORCID:** 0000-0002-5737-3487

---

[1] The opinions expressed in this publication are those of the author. They do not purport to reflect the opinions or views of her employer.